\documentclass[twocolumn,showpacs,preprintnumbers,amsmath,amssymb]{revtex4}

\usepackage{graphicx}
\usepackage{dcolumn}
\usepackage{bm}

\begin{document}


\title{Phase transition behavior in a cellular automaton model with different initial configurations}


\author{Wei Zhang$^{1,2}$}
 \email{tzwphys@jnu.edu.cn; wzhang2007065@gmail.com}
\author{Wei Zhang$^{2}$}
 \email{twzhang@jnu.edu.cn}
\author{Wei Chen$^{1}$}%
 \email{phchenwei@gmail.com}
\affiliation{%
$^1$Department of Physics, Fudan University, Shanghai 200433, China\\
$^2$Department of Physics, Jinan University, Guangzhou 510632, China}

\date{\today}

\begin{abstract}
We investigate the dynamical transition from free-flow to jammed traffic, which is related to the divergence of the relaxation time and susceptibility of the energy dissipation rate $E_d$, in the Nagel-Schreckenberg (NS) model with two different initial configurations. Different initial configurations give rise to distinct phase transition. We argue that the phase transition of the deterministic NS model with megajam and random initial configuration is first- and second-order phase transition, respectively. The energy dissipation rate $E_d$ and relaxation time follow power-law behavior in some cases. The associated dynamic exponents have also been presented.
\end{abstract}

\maketitle

\section{\label{sec:level1}INTRODUCTION}

Nonequilibrium phase transitions and various nonlinear dynamical phenomena in traffic system have attracted much attention of a community of physicists in recent years. There are many theories to describe traffic phenomenon such as fluid-dynamical theories, kinetic theories, car-following theories and cellular automata (CA) models\cite{1,2,3}. The advantages of the cellular automata approaches show the flexibility to adapt complicated features observed in real traffic \cite{1,4}. In addition , CA theory is also a simple and useful approach for the study of nonequilibrium steady states and their transition mechanisms. So CA theory has been extensively applied and investigated. The Nagel-Schreckenberg (NS) model is a basic CA models describing one-lane traffic flow and phase transition\cite{5}.
	
The question of the dynamical transition from free-flow phase to jammed phase in NS model has been investigated by several scholars\cite{6,7,8,9,10,11,12,13,14}. However, to our knowledge, the effects of the initial configuration on phase transition behavior have not been explored in detail so far, and should be further investigated
	
The energy dissipation rate $E_d$ proposed by us in Refs.[15] is related to traffic phase transition, and can be viewed as an order parameter. In the deterministic NS model, there is a critical density below which the parameter is zero which is associated with a free-flow phase, but over which the order parameter is not zero anymore representing the jammed phase. The study of energy dissipation in traffic system has important realistic significance. According to Refs.[16], more than 20\% fuel consumption and air pollution is caused by impeded and ''go and stop'' traffic. Due to the relevance of this parameter for realistic case it is important to understand it's phase transition behavior thoroughly.
	
In this paper, using the order parameter $E_d,$ we study the nonequilibrium phase transition in the NS model with random and megajam initial configurations. We argue that the phase transition in the deterministic NS model with megajam and random initial configuration is first- and second-order phase transition, respectively. The relaxation time of $E_d$ in deterministic NS model with the two initial configuration are distinct and will be analyzed. And an associated susceptibility is numerically studied. Some critical exponents will be presented in the following section.
	
The paper is organized as follows. Section II is devoted to the description of the NS model and the definition of energy dissipation rate. In section III, the numerical studies of relaxation time and susceptibility of energy dissipation rate in NS models are given, and the influences of the initial configuration on phase transition are considered. The results are summarized in section IV.

\section{DESCRIPTION OF THE MODEL AND ORDER PARAMETER}

The model is defined on a single lane road consisting of $L$ cells of equal size numbered by $i=1,$ $2,$ $\cdots ,$ $L$ and the time is discrete. Each site can be either empty or occupied by a car with the speed $v=0,$ $1,$ $2,\cdots $ $,$ $v_{\max }$, where $v_{\max }$ is the speed limit. Let $x(i,t)$ and $v(i,t)$ denote the position and the velocity of the $i$th car at time $t$, respectively. The number of empty cells in front of the $i$th car is denoted by $d(i,t)=x(i+1,t)-x(i,t)-1$. The following four steps for all cars update in parallel with periodic boundary.

(1) Acceleration:

$v(i,t+1/3)\rightarrow \min [v(i,t)+1,v_{\max }];$

(2) Slowing down:

$v(i,t+2/3)\rightarrow \min [v(i,t+1/3),d(i,t)];$

(3) Stochastic braking:

$v(i,t+1)\rightarrow \max [v(i,t+2/3)-1,0]$ with the probability
$p;$

(4) Movement: $x(i,t+1)\rightarrow x(i,t)+v(i,t+1).$

Iteration over these four update rules already gives realistic results such as the spontaneous formation of traffic jams, ''go and stop'' wave. With increasing vehicle density dynamitic transition from free-flow phase to jammed state occurs. There are two types of initial configuration: random condition where all vehicles' positions are distributed randomly, and megajam configuration where all vehicles stand in one big cluster.
	
The kinetic energy of the car moving with the velocity $v$ is $mv^2/2$, where $m$ is the mass of the vehicle. When braking the kinetic energy reduces. Let $E_d$ denotes energy dissipation rate per time step per vehicle. For simple, we neglect rolling and air drag dissipation and other dissipation such as the energy needed to keep the motor running while the vehicle is standing and moving in our analysis, i.e. we only consider the energy lost caused by speed-down. The dissipated energy of $i$th car from time $t-1$ to $t$ is defined by

\[
e(i,t)=%
{\frac m2\left[ v^2(i,t-1)-v^2(i,t)\right] \quad \text{for }v(i,t)<v(i,t-1) \atopwithdelims\{. 0\qquad \qquad \qquad \qquad \qquad ~~\text{for }v(i,t)\geqslant v(i,t-1).}%
\qquad \left( 1\right)
\]
Thus, the energy dissipation rate

\[
E_d=\frac 1T\frac 1N\sum_{t=t_0+1}^{t_0+T}\sum_{i=1}^Ne(i,t),\qquad
\left( 2\right)
\]
where $N$ is the number of vehicles in the system and $t_0$ is the relaxation time, taken as $t_0=10L$ unless stated otherwise. Assuming that energy dissipation per vehicle at time $t$ is $e(t)=\frac 1N\sum_{i=1}^Ne(i,t).$ We consider the relaxation time $\tau $ of $e(t)$ need to get to the steady state as the time obtained when $\left| E_d-e(t)\right| <10^{-4}.$
	
In this model, the particles are ''self-driven'' and the kinetic energy increases in the acceleration step. In the stationary state, the value of the increased energy while accelerating is equivalent to that of the dissipated energy caused by speed-down, and the kinetic energy is constant in the system. Generally, the mean density is denoted by $\rho =N/L.$

\begin{figure}
\includegraphics[height=6cm]{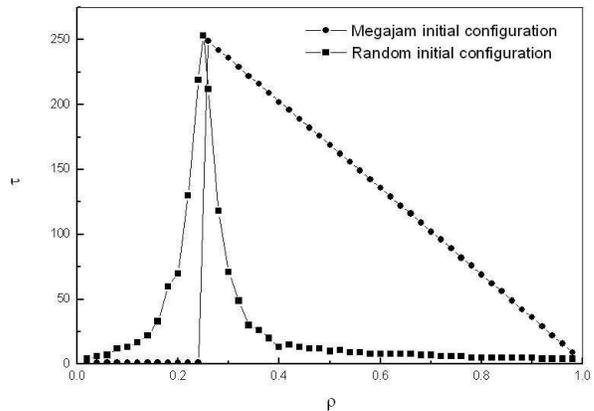}
\caption{\label{fig:epsart} The relaxation time $\tau $ as a function of the vehicle density $\rho $ in the deterministic NS model with different initial configuration for the case of $v_{\max }=3$, $L=1000$.}
\end{figure}

\section{NUMERICAL RESULTS}

First, we investigate the influences of the initial configuration on the
relaxation time $\tau $ of energy dissipation $e(t)$ in the deterministic NS
model. In the deterministic case, the stochastic braking is not considered,
i.e. $p=0$. Figure 1 shows the relaxation time $\tau $ as a function of the
vehicle density $\rho $ with megajam and random initial configuration in the
case of $v_{\max }=3,L=1000$. As shown in Fig. 1, there is a critical
slowing down; and the relaxation to the steady state becomes quite slow
close to the critical density $\rho _c=1/(1+v_{\max })$. The relaxation time
$\tau $ diverges at the critical density $\rho _c$ in the model with random
initial configuration, which are consistent with a second-order phase
transition. In the case of megajam initial configuration, however, the
relaxation time $\tau $ is invalid below the critical density $\rho _c$ for
there is no energy dissipation occurs at any time. Above the critical
density, $\tau $ decreases linearly with increasing the vehicle density. As
reported in Refs.[15 ], above the critical density $\rho _c,$ energy
dissipation rate $E_d$ occurs abruptly and reaches the maximum value in the
system with megajam initial condition, which is different from that with
random initial configuration. The phase transition is not continuous. Thus,
we argue that the dynamical transition from free-flow traffic to jammed
state in the system with megajam initial condition can be viewed as
first-order phase transition. In the jammed phase, the relaxation to the
steady state in the system with megajam initial condition is slower than
that with random initial configuration, as shown in Fig. 1.

\begin{figure}
\includegraphics[height=6cm]{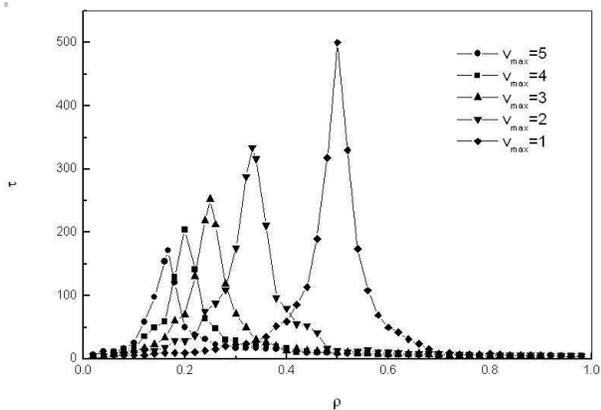}
\caption{\label{fig:epsart} The relaxation time $\tau $ as a function of the vehicle density $\rho $ for various values of the speed limit $v_{\max }$ in the deterministic NS model with random initial configuration for the case of $L=1000$.}
\end{figure}

Figure 2 shows the relaxation time $\tau $ as a function of the vehicle
density $\rho $ with different values of the speed limit $v_{\max }$ in the
case of random initial configuration. From figure 2, we see that the maximal
value of relaxation time $\tau _m$ occurs at the critical density $\rho _c.$
And the time $\tau _m$ decreases with the increase of speed limit $v_{\max }.
$ The relaxation time $\tau _m$ is given as

\[
\tau _m=\frac L{v_{\max }+1},\qquad \qquad (3)
\]
which is compatible with the results obtained with another order parameter%
\cite{14}. Figure 3 shows the relaxation time $\tau _m$ as a function of the
system size $L$ with different values of the speed limit $v_{\max }.$ As
shown in figure 3, symbol data are obtained from computer simulations, and
solid lines correspond to analytic results of the formula (3).

\begin{figure}
\includegraphics[height=6cm]{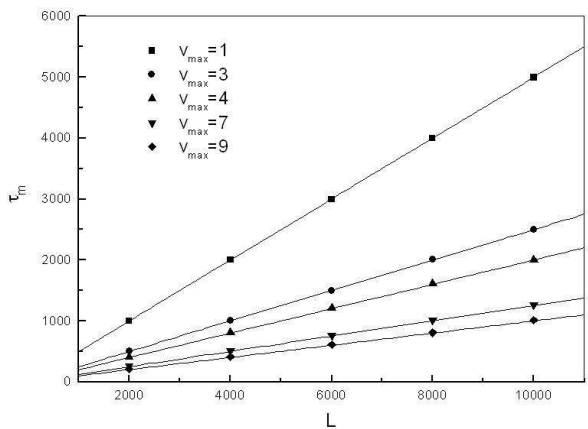}
\caption{\label{fig:epsart} The relaxation time $\tau _m$ as a function of the lattice size $L$ for various values of the speed limit $v_{\max }$ in the deterministic NS model with random initial configuration. Symbol data are obtained from computer simulations, and solid line corresponds to analytic results of the formula (3).}
\end{figure}

Figure 4 shows the relaxation time $\tau $ as a function of the vehicle
density $\rho $ with different values of the speed limit $v_{\max }$ in the
case of megajam initial configuration. From figure 4, we see that the
relaxation time $\tau $  decreases linearly with the increase of the vehicle
density over the critical density $\rho _c$. The relaxation time $\tau $ can
be written as

\[
\tau =\frac L{v_{\max }}(1-\rho ).\qquad \qquad (4)
\]
In figure 4, symbol data are obtained from computer simulations, and solid
lines correspond to analytic results of the formula (4). Compared with the
theoretical results and simulation data, excellent agreement can be obtained.

\begin{figure}
\includegraphics[height=6cm]{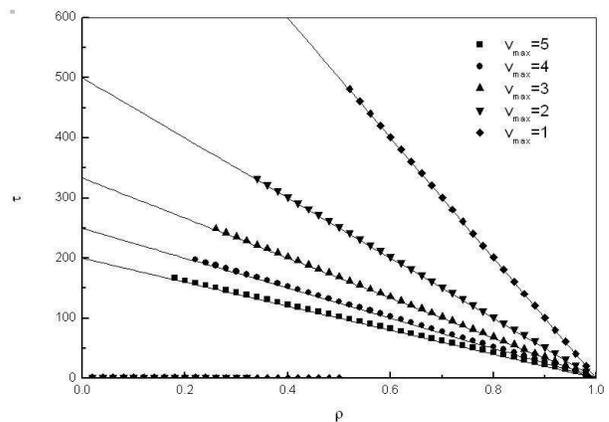}
\caption{\label{fig:epsart} The relaxation time $\tau $ as a function of the vehicle density $\rho $ for various values of the speed limit $v_{\max }$ in the deterministic NS model with megajam initial configuration for the case of $L=1000$.. Symbol data are obtained from computer simulations, and solid line corresponds to analytic results of the formula (4).}
\end{figure}

In the deterministic case, i.e. $p=0$, the order parameter $E_d$ is zero
when $\rho \leq \rho _c$ and $E_d>0$ in the density interval $1>\rho >\rho
_c.$ However, in the case of $p\neq 0$, the parameter $E_d$ is not zero
anymore. The phase transition observed in the deterministic case is
destroyed by the stochastic braking probability $p$. The probability $p$ is
the conjugated parameter of the order parameter of NS model. In order to
analyze the relationship between the probability $p,$ energy dissipation
rate $E_d$ and phase behavior, we define the associated susceptibility

\[
\chi _p=\left. \frac{\partial E_d}{\partial p}\right| _{p=0}.\qquad \qquad
(5)
\]

\begin{figure}
\includegraphics[height=6cm]{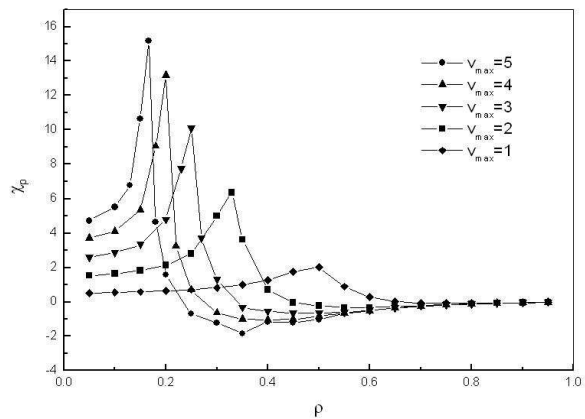}
\caption{\label{fig:epsart} The susceptibility $\chi _p$ as a function of the vehicle density $\rho $ in NS model with random initial configuration for various values of the speed limit $v_{\max },$ for the case of $L=10000$ and $p=0.01.$}
\end{figure}

\begin{figure}
\includegraphics[height=6cm]{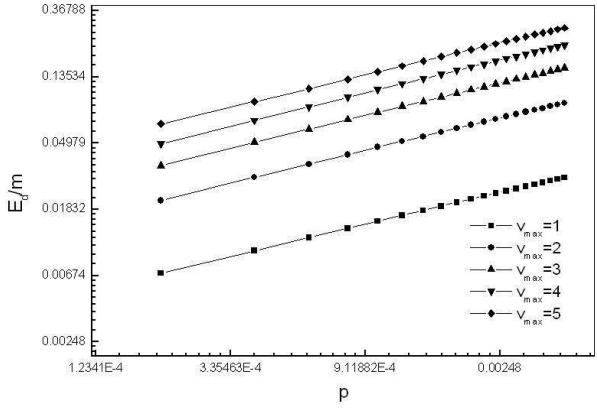}
\caption{\label{fig:epsart} The log-log plot of energy dissipation rate $E_d$ (scaled by $m$ ) at the critical density $\rho _c$ as a function of the stochastic braking probability $p$ in NS model with random initial configuration for the case of $L=10000$, for various values of the speed limit $v_{\max }$.}
\end{figure}

Figure 5 exhibits the relation of $\chi _p$ to the vehicle density $\rho $
with various values of the speed limit and random initial condition in the
case of $p=0.01,L=10000$. As shown in figure 5, the susceptibility $\chi _p$
first increases with the density $\rho $, then it decreases with $\rho $
above the critical density $\rho _c$ where a maximum value is reached. At
high density region, the susceptibility $\chi _p$ is negative, i.e. energy
dissipation rate $E_d$ decreases with increasing the probability $p$. When $%
\rho >0.8,$ the curves converges into one curve and $v_{\max }$ has no
influence on $\chi _p$. The peak's values increase with the increase of the
speed limit $v_{\max }.$ The peak of $\chi _p$ tends to diverge at the
critical density $\rho _c$ in the case of $p\rightarrow 0$. The divergence
of $\chi _p$ is relevant to the traffic phase transition.

From figure 5, we can observe that $\chi _p$ $\propto v_{\max }$ when $\rho
\rightarrow 0$. At the critical density $\rho _c,$ for $p\rightarrow 0,$ we
find that the order parameter $E_d{}$ follows a power-law behavior of the
form

\[
E_d{}\propto p^\gamma .\qquad \qquad (6)
\]

\begin{figure}
\includegraphics[height=6cm]{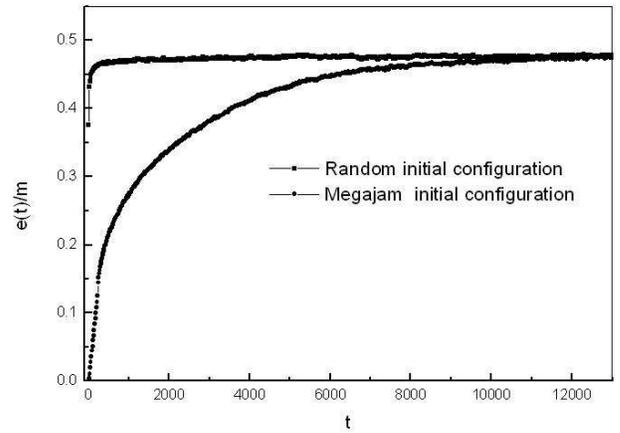}
\caption{\label{fig:epsart} Time evolutions of energy dissipation $e(t)$ (scaled by $m$ ) starting from the random and megajam initial configuration for the case of $v_{\max }=3$, $p=0.005$, $\rho =0.3$ and $L=1000$.}
\end{figure}

\begin{figure}
\includegraphics[height=6cm]{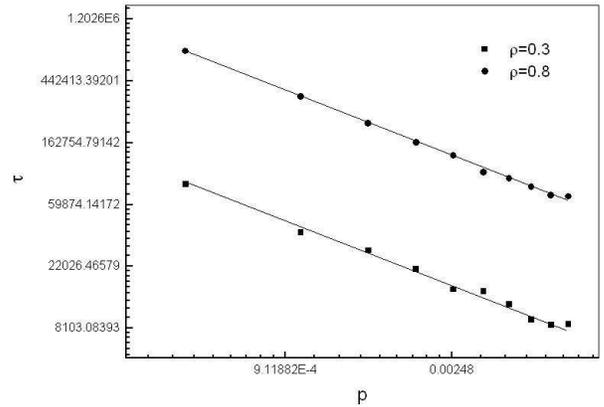}
\caption{\label{fig:epsart} The relaxation time $\tau $ as a function of stochastic braking probability $p$ near the limit $p\rightarrow 0$ for several fixed density in NS model with megajam initial configuration for the case of $v_{\max }=3$ and $L=1000$. Lines are guides for the eyes.}
\end{figure}

Figure 6 shows that the critical exponent $\gamma \approx 0.483\pm 0.005$
and the speed limit has no influences on $\gamma .$ In the NS model with
megajam initial configuration, the behavior of $\chi _p$ is similar with
that with random initial condition. The dynamic exponent $\gamma $ remains
unchanged when varying the initial configuration. However, the initial
condition has effects on the relaxation time $\tau $ near the limit $%
p\rightarrow 0$. Figure 7 exhibits time evolutions of $e(t)$ starting from
the megajam and random initial configuration in the case of $v_{\max
}=3,p=0.005,L=1000$ and $\rho =0.3.$ As shown in figure 7, the relaxation to
steady state in the system with megajam initial condition is slower than
that with random configuration, even if the values of $e(t)$ at the steady
state are unique. Figure 8 exhibits the relation of relaxation time $\tau $
to the stochastic braking probability $p$ near the limit $p\rightarrow 0$
with various values of vehicle density. From figure 8, we see that

\[
\tau \propto p^{-\delta },\qquad \qquad (7)
\]
and the dynamic exponent $\delta \approx 1.04\pm 0.02.$

\begin{figure}
\includegraphics[height=6cm]{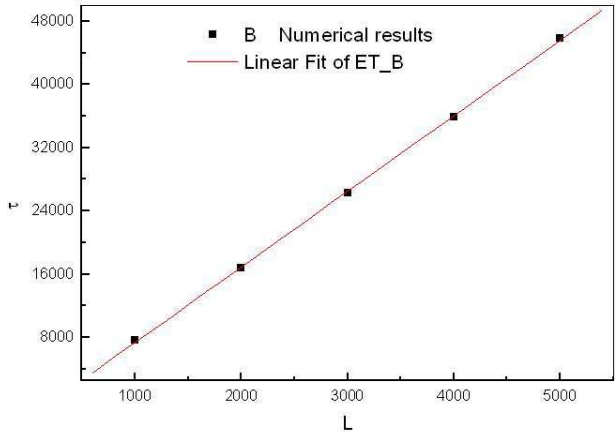}
\caption{\label{fig:epsart} The relaxation time $\tau $ as a function of the system size $L$ with megajam initial configuration for the case of $v_{\max }=3$,$\rho =0.3$ and $p=0.005$.}
\end{figure}

Figure 9 shows the relaxation time $\tau $ as a function of the system
size $L$ with megajam initial configuration in the
case of $v_{\max }=3$,$\rho =0.3$ and $p=0.005$. As shown in figure 9, the relaxation time $\tau $ has a scaling form

\[
\tau \propto L^{\nu},\qquad \qquad (8)
\]
and the exponent $\nu \approx 1,$ which is compatible with the results presented in Refs.[17,18].

\section{SUMMARY}

In this paper, we investigate traffic phase transition behavior in the NS
model considering different initial configurations. Different initial
condition gives rise to distinct phase transition. We argue that the phase
transition in the system with megajam and random configuration is first- and
second-order phase transition, respectively. Using the order parameter $E_d$%
, we numerically studied the relaxation time $\tau $ and susceptibility $%
\chi _p.$ The two quantities diverge at the critical density $\rho _c.$ We
analyzed the relaxation time $\tau _m$ and $\tau $ theoretically.
Theoretical analyses give an excellent agreement with numerical results.
Near the limit $p\rightarrow 0$, the parameter $E_d$ and relaxation time $%
\tau $ follow a power-law behavior. The dynamic exponent $\gamma \approx
0.483\pm 0.005$, $\delta \approx 1.04\pm 0.02$ and $\nu \approx 1.$

When $p\nrightarrow0$, the phase transition behavior becomes more complicated. We will
make further investigation using the order parameter $E_d$. The associated
dynamic exponents and scaling laws will be presented elsewhere.

\end{document}